\documentclass[aps,prl,twocolumn,groupedaddress,showpacs]{revtex4}

\usepackage{graphicx}

\begin{document}

\title{Nonclassical Properties of Coherent States}

\author{Lars M. Johansen}

\affiliation{Department of Technology, Buskerud University
College, P.O. Box 251, N-3601 Kongsberg, Norway}

\email{lars.m.johansen@hibu.no}

\date{\today}

\begin{abstract}

It is demonstrated that a weak measurement of the squared
quadrature observable may yield negative values for coherent
states. This result cannot be reproduced by a classical theory
where quadratures are stochastic $c$-numbers. The real part of the
weak value is a conditional moment of the Margenau-Hill
distribution. The nonclassicality of coherent states can be
associated with negative values of the Margenau-Hill distribution.
A more general type of weak measurement is considered, where the
pointer can be in an arbitrary state, pure or mixed.

\end{abstract}


\maketitle

Harmonic oscillator coherent states were first investigated by
Schr{\"o}dinger, who was looking for classical-like states
\cite{Schroedinger-stetUEbeMikrMakr:26}. There are several ways in
which coherent states are the ``most classical" of any pure state.
They keep their shape, not spreading out as they move in the
harmonic oscillator potential
\cite{Schroedinger-stetUEbeMikrMakr:26}. They minimize
Heisenberg's uncertainty relation, with equal uncertainty in both
quadratures. In this way, they are the closest possible quantum
mechanical representation of a point in phase space. The term
``coherent state" was introduced by Glauber
\cite{Glauber-CoheIncoStatRadi:63}. He demonstrated that coherent
states are produced when an essentially classical current
interacts with the radiation field
\cite{Glauber-CoheIncoStatRadi:63}. Aharonov \emph{et. al.}
demonstrated that coherent states are the only pure states that
produce independent output when split in two
\cite{Aharonov+FalkoffETAL-QuanCharClasRadi:66}. Zurek \emph{et.
al.} have demonstrated that coherent states are natural ``pointer
states" for a harmonic oscillator weakly coupled to a thermal
environment \cite{Zurek+HabibETAL-CoheStatDeco:93}.

Glauber and Sudarshan demonstrated that any density operator can
be expanded in terms of coherent states
\cite{Glauber-CoheIncoStatRadi:63,Sudarshan-EquiSemiQuanMech:63}
\begin{equation}
    \hat{\rho} = \int d^2 \alpha P(\alpha) \mid \alpha \rangle
    \langle \alpha \mid.
\end{equation}
The weight function $P(\alpha)$ is known as the $P$-distribution.
Glauber defined nonclassical states as those for which the
$P$-distribution fails to be a probability density. More
specifically, nonclassical states have a $P$-distribution which is
negative or more singular than a $\delta$-function
\cite{Glauber-CoheIncoStatRadi:63,%
Titulaer+Glauber-CorrFuncCoheFiel:65,%
Hillery-Claspurestatcohe:85,%
Mandel+Wolf-OptiCoheQuanOpti:95,%
Dodonov-Noncstatquanopti:02}. This criterion is the basis of
various measures of ``nonclassicality"
\cite{Hillery-Noncdistquanopti:87,%
Lee-Measnoncnoncstat:91,%
Vogel-NoncStat:00,%
Richter+Vogel-NoncQuanStat:02}.

It is the purpose of this Letter to demonstrate that a quantum
state may be nonclassical even though the $P$-distribution is a
probability density, and that also coherent states display
nonclassical characteristics. In this Letter, we associate
nonclassicality with the failure of the Margenau-Hill distribution
\cite{Margenau+Hill-CorrbetwMeasQuan:61} to be a probability
distribution. The Margenau-Hill distribution yields correct
marginal distributions, just as the Wigner distribution
\cite{Wigner-QuanCorrTherEqui:32}. But in contrast to the Wigner
distribution, it is negative for coherent states
\cite{Praxmeyer+Wodkiewicz-QuanInteKirkrepr:03}. We give an
operational significance to conditional moments of the
Margenau-Hill distribution by demonstrating that they can be
observed in ``weak measurements". Weak measurements were proposed
by Aharonov \emph{et. al.}
\cite{Aharonov+AlbertETAL-ResuMeasCompSpin:88}. Their suggestion
was initially met with criticism
\cite{Leggett-CommResuMeasComp:89,Peres-QuanMeaswithPost:89,%
Aharonov+Vaidman-AharVaidRepl:89}, but has since been confirmed
in various ways (see, e.g., \cite{Duck+StevensonETAL-senswhicweakmeas:89,%
Ritchie+StoryETAL-RealMeasWeakValu:91,%
Hulet+RitchieETAL-MeasWeakValu:97,%
Parks+CullinETAL-ObsemeasoptiAhar:98}). The results reported in
this Letter are related to a paper by Aharonov {\em et al.}
\cite{Aharonov+PopescuETAL-MeasErroNegaKine:93}, which
demonstrated that a weak measurement of kinetic energy of a
particle in a classically forbidden region might yield negative
values.

In the original von Neumann measurement scheme
\cite{Neumann-MathFounQuanMech:55}, it was found that in order to
distinguish different eigenvalues of the object, the pointer
should be in a state with small uncertainty in the pointer
position. Aharonov \emph{et. al.}.
\cite{Aharonov+AlbertETAL-ResuMeasCompSpin:88} proposed to define
weak measurements by using a pointer with a large pointer position
uncertainty. In this Letter, we abandon this condition. Instead,
we assume that the interaction between the pointer and the object
is sufficiently weak. Thus, the pointer can be in an arbitrary
state, pure or mixed. We impose only one condition on the pointer,
namely that the current density should vanish.

We consider an object and a pointer described by the density
operators $\hat{\rho}_s$ and $\hat{\rho}_a$, respectively. Prior
to the measurement interaction, the combined object plus pointer
is assumed to be in a product state $\hat{\rho}_0 = \hat{\rho}_s
\otimes \hat{\rho}_a$. We wish to perform a weak measurement of an
arbitrary object observable $\hat{c}$. To this end, we shall
assume that the interaction part of the Hamiltonian has the form
\begin{equation}\label{eq:quantuminteraction}
    \hat{H}_\epsilon = \epsilon \delta(t) \; \hat{c} \otimes \hat{P}.
\end{equation}
This interaction Hamiltonian is essentially the same as proposed
in Ref. \cite{Aharonov+AlbertETAL-ResuMeasCompSpin:88}, except
that we have introduced an interaction strength $\epsilon$. It is
a generalization of the interaction Hamiltonian proposed by von
Neumann \cite{Neumann-MathFounQuanMech:55}. It has been discussed
in detail for ``strong" measurements in Ref.
\cite{Haake+Walls-Overamplmetequan:87}. $\hat{P}$ is the momentum
observable of the pointer. We will consistently denote observables
associated with the pointer by capital letters. We assume that
during the measurement interaction, the interaction part of the
Hamiltonian dominates the time evolution. Nevertheless, we shall
assume that the interaction between the object and pointer is
weak, i.e., $\epsilon$ is so small that we can perform a series
expansion to first order in $\epsilon$. The possibility of
realizing this or similar interactions experimentally will be
discussed at the end of this Letter.

Because of the interaction between the object and pointer, the
density operator evolves to $\hat{\rho}_\epsilon =
\hat{U}_\epsilon \hat{\rho}_0 \hat{U}_\epsilon^{\dag}$, where the
unitary evolution operator $\hat{U}_\epsilon$ is (setting
$\hbar=1$)
\begin{equation}
    \hat{U}_\epsilon = e^{- i \int \hat{H}_\epsilon (t) dt } =
    e^{- i \epsilon \hat{c} \otimes \hat{P}}.
\end{equation}
In this experiment, we are interested in the final values of the
pointer position $\hat{Q}$ and the object position $\hat{q}$. The
joint probability distribution for these observables is
\begin{equation}
    \rho_\epsilon(Q,q) = \langle q \mid \otimes \langle Q \mid
    \hat{\rho}_\epsilon \mid Q \rangle \otimes \mid q \rangle.
\end{equation}
We require that the current density of the pointer vanishes,
\begin{equation}
    \langle Q \mid \hat{P} \hat{\rho}_a \mid Q \rangle +
    \langle Q \mid \hat{\rho}_a \hat{P} \mid Q \rangle = 0.
\end{equation}
This is our only restriction on the state of the pointer. It can
then be shown that to the first order in $\epsilon$, the
probability density for the pointer position $Q$ conditioned on
the object position $q$ reads \cite{Johansen-WeakMeaswithArbi:04}
\begin{equation}
    \rho_\epsilon(Q \mid q) = {\rho_\epsilon(Q,q) \over \int dQ
    \rho_\epsilon(Q,q)} \approx \hat{\cal T} \langle Q \mid
    \hat{\rho}_a \mid Q \rangle,
\end{equation}
where
\begin{equation}
    \hat{\cal T} = 1 - \epsilon \mathrm{Re}(c_w) {\partial
    \over \partial Q}
\end{equation}
is a first order translation operator, and where
\begin{equation}\label{eq:mixedweak}
    c_w(q) = {\langle q \mid \hat{c} \hat{\rho}_s \mid q
    \rangle \over \langle q \mid \hat{\rho}_s \mid q
    \rangle}
\end{equation}
is the weak value of $\hat{c}$ for an object preselected in a
mixed state $\hat{\rho}_s$ and postselected in the eigenstate
$\mid q \rangle$
\cite{Aharonov+AlbertETAL-ResuMeasCompSpin:88,%
Hall-Exacuncerela:01,Johansen-Whatvaluobsebetw:03}. This shows
that the pointer position $Q$ for a given object position $q$ has
been translated by a distance $\epsilon \mathrm{Re} (c_w)$. The
basic condition for a weak measurement is that the translation of
the pointer should be small compared to the standard deviation of
the pointer $\sigma$, i.e. $\mid \epsilon \mathrm{Re} \left ( c_w
\right ) \mid \ll \sigma$.

\begin{figure}
\includegraphics[width=6cm]{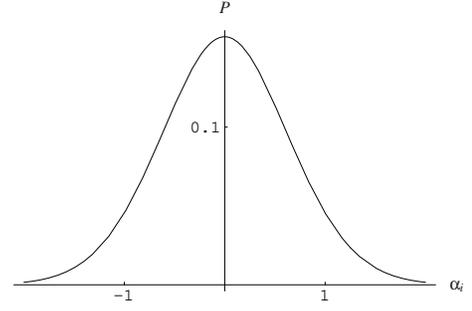}
\caption{The probability of observing a negative weak value for a
coherent state with amplitude $(\alpha_r + i \alpha_i)/\sqrt{2}$.
It is plotted as a function of the imaginary component $\alpha_i$.
The probability is independent of the real component $\alpha_r$.}
\label{fig:prob}
\end{figure}

We now demonstrate that a weak measurement of the positive
operator $\hat{p}^2$ may yield negative values for a coherent
state. Consider the quadrature representation of a coherent state
$\mid \alpha \rangle$ (with $\omega=1$)
\cite{Louisell-QuanStatPropRadi:73}
\begin{equation}
    \label{eq:coherent}
    \langle q \mid \alpha \rangle = \pi^{-1/4} \exp \left [ -{q^2
    \over 2} + \sqrt{2} \, \alpha \, q - {1 \over 2} \mid \alpha \mid^2
    - {1 \over 2} \, \alpha^2 \right ].
\end{equation}
The weak value of $\hat{p}^2$ for an ensemble preselected in the
coherent state $\mid \alpha \rangle$ and postselected in the
quadrature eigenstate $\mid q \rangle$ then is
\begin{equation}
    (p^2)_w = {- \partial^2 \langle q \mid \alpha \rangle /
    \partial q^2 \over \langle q \mid \alpha \rangle} =
    1- (q-\sqrt{2} \alpha)^2.
\end{equation}
The real part of the weak value is $\mathrm{Re}[(p^2)_w] = 1 +
\alpha_i^2 - (q- \alpha_r)^2$, where we have introduced the
notation $\alpha = (\alpha_r + i \alpha_i)/\sqrt{2}$. We see that
$\mathrm{Re}[(p^2)_w]$ is negative if $(q - \alpha_r)^2 > 1 +
\alpha_i^2$. The surprising conclusion is that the weak value of
$\hat{p}^2$ can be negative for coherent states, although
$\hat{p}^2$ has only nonnegative eigenvalues. The probability of
obtaining a negative value is
\begin{eqnarray}
    P &=& \int_{-\infty}^{\alpha_r - \sqrt{1+\alpha_i^2}} \mid
    \langle q \mid \alpha \rangle \mid^2 dq \nonumber \\
    &+& \int_{\alpha_r + \sqrt{1+\alpha_i^2}}^\infty
    \mid \langle q \mid \alpha \rangle \mid^2 dq.
\end{eqnarray}
This is found to be $\mathrm{erfc} \sqrt{1+\alpha_i^2}$, where
$\mathrm{erfc}(x)$ is the complementary error function. This
function is plotted in Fig. \ref{fig:prob}. It has a maximum when
the imaginary part of the coherent state amplitude vanishes, in
which case it equals $\mathrm{erfc}(1) \approx 0.16$.

We now demonstrate that a negative $\mathrm{Re}[(p^2)_w]$ is
closely related to negativity of the Margenau-Hill distribution.
Consider the weak value as defined in Eq. (\ref{eq:mixedweak}) for
the observable $\hat{c} = \hat{p}^n$. By inserting the
completeness relation $\int dp \mid p \rangle \langle p \mid = 1$
in the numerator, we find that the weak value of $\hat{p}^n$ can
be written as
\begin{equation}
    (p^n)_w = {\int dp \; p^n \; S(q,p) \over \langle q \mid
    \hat{\rho}_s \mid q \rangle},
\end{equation}
where
\begin{equation}\label{eq:standard}
    S(q,p) = \langle q \mid p \rangle \langle p \mid
    \hat{\rho}_s \mid q \rangle
\end{equation}
is the standard ordered distribution
\cite{Mehta-PhasFormDynaCano:64}. This is the complex conjugate of
the Kirkwood distribution \cite{Kirkwood-QuanStatAlmoClas:33}. The
Margenau-Hill distribution is the real part of the standard
ordered or Kirkwood distributions, so that
\begin{equation}
    \mathrm{Re}[(p^n)_w] = {\int dp \; p^n \; M(q,p) \over \langle
    q \mid \hat{\rho}_s \mid q \rangle},
\end{equation}
where $M(q,p)$ is the Margenau-Hill distribution.
$\mathrm{Re}[(p^n)_w]$ is a conditional moment of the
Margenau-Hill distribution ( see also
\cite{Sutherland-Joindistindiwave:82,
Steinberg-MuchTimeDoesTunn:95,Steinberg-Condprobquantheo:95,
Luis-Phasdistclascomp:03}). Clearly, $\mathrm{Re}[(p^2)_w]$ cannot
be negative unless the Margenau-Hill distribution is negative.

\begin{figure}
\includegraphics[width=7cm]{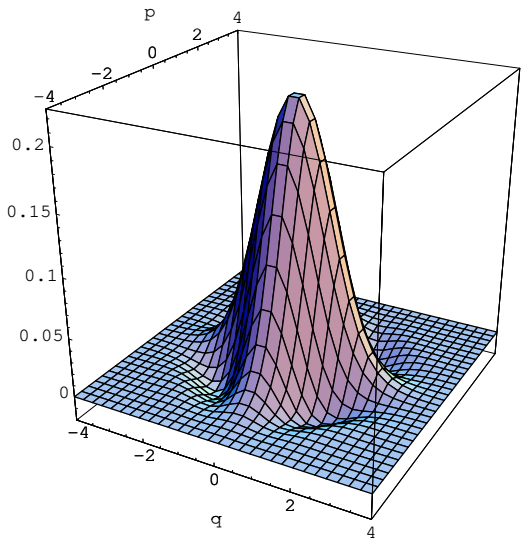}
\caption{The Margenau-Hill distribution for vacuum. It is negative
in a certain domain. The Margenau-Hill distribution for an
arbitrary coherent state is just a displaced vacuum state.}
\label{fig:mh}
\end{figure}

Combining Eqs. (\ref{eq:coherent}) and (\ref{eq:standard}), the
standard ordered distribution for a coherent state $\mid \alpha
\rangle$ is easily found. The standard ordered distribution for
vacuum can be written as
\begin{equation}
    S_0(q,p) = {1 \over \sqrt{2} \pi} e^{- {1 \over 2}
    (q^2 + p^2) + i q p}.
\end{equation}
The standard ordered distribution for a coherent state may then be
expressed in terms of the displaced vacuum, $S_\alpha(q,p) =
S_0(q-\alpha_r,p-\alpha_i)$. The real part of $S_0$, the
Margenau-Hill distribution, has been plotted in Fig. \ref{fig:mh}
(see also Ref. \cite{Praxmeyer+Wodkiewicz-QuanInteKirkrepr:03}).
It clearly has negative regions. This is the reason why a weak
measurements of the positive operator $\hat{p}^2$ may yield
negative values for a coherent state.

A classical system cannot reproduce the negative weak values that
we have found. To demonstrate this, assume that the object and
pointer both can be described by a classical phase space
distribution. Prior to the measurement interaction, we assume that
the object plus pointer are in a product state $F_0 = F_s(q,p)
F_a(Q,P)$, where again capital letters denote the pointer. We
consider a weak measurement of a general classical object variable
$c(q,p)$, and assume that the interaction Hamiltonian is
\begin{equation}\label{eq:classicalinteraction}
    H_\epsilon = \epsilon \, \delta(t) \, c(q,p) \, P.
\end{equation}
This is the classical equivalent of the quantum interaction term
(\ref{eq:quantuminteraction}). We again assume that the
interaction Hamiltonian dominates over any other terms in the
Hamiltonian during the short time of interaction. The equation of
motion is given by the classical Liouville theorem,
\begin{equation}
    {\partial F \over \partial t} =-\{ F, H_\epsilon \}.
\end{equation}
Due to the interaction, the joint phase space distribution evolves
to $F_\epsilon(q,p,Q,P)$. The joint probability density for the
two position variables then reads
\begin{equation}
    \rho_\epsilon(Q, q) = \int dp \int d P F_\epsilon(q,p,Q,P).
\end{equation}
By assuming once more that the current density of the pointer
vanishes,
\begin{equation}
    \int dP \; P \; F_a(Q,P) = 0,
\end{equation}
it can be shown that to the first order in $\epsilon$, the
probability density for the pointer position $Q$ conditioned on
the object position $q$ is \cite{Johansen-WeakMeaswithArbi:04}
\begin{equation}
    \rho_\epsilon(Q \mid q) = {\rho_\epsilon(Q, q) \over \int dQ
    \rho_\epsilon(Q, q)} \approx \hat{\cal T}_c f_a(Q),
\end{equation}
where
\begin{equation}
    f_a(Q) = \int dp F_a(Q,P)
\end{equation}
and
\begin{equation}
    \hat{\cal T}_c = 1 - \epsilon c_w {\partial \over \partial Q}
\end{equation}
is a first order translation operator. Here we have introduced
\begin{equation}\label{eq:classicalweakvalue}
    c_w = {\int dp \; c(q,p) \; F_s(q,p),\\
    \over \int dp \, F_s(q,p)}.
\end{equation}
which is the ``classical weak value" of $c(q,p)$. We see that the
classical weak value is the conditional expectation value of that
variable. In other words, $c_w$ is simply the expectation value of
$c(q,p)$ ``given" $q$. This shows that the pointer $Q$ has been
translated by a distance $\epsilon c_w$. In this case, the
measurement is weak provided that $\mid \epsilon c_w \mid \ll
\sigma$, where $\sigma$ is the standard deviation of the pointer
position.

It follows straightforwardly from Eq.
(\ref{eq:classicalweakvalue}) that if $c(q,p) \ge 0$, then due to
a nonnegative integrand, $c_w \ge 0$. The classical weak value of
a positive observable cannot be negative. However, we have just
seen that this condition can be violated for positive observables
on coherent states. We therefore conclude that coherent states
possess nonclassical properties.

Our analysis assumed an interaction of the form
(\ref{eq:quantuminteraction}). However, just as a standard,
projective von Neumann measurement is not dependent on the
specific interaction Hamiltonian proposed by von Neumann
\cite{Neumann-MathFounQuanMech:55}, it is not to be expected that
weak measurements are critically dependent on the specific form of
interaction Hamiltonian proposed here. For a massive particle in a
harmonic oscillator potential, it should be possible to employ
almost any measurement scheme for kinetic energy provided that the
interaction between the particle and the pointer is sufficiently
weak. The position measurement can be provided by using a detector
with spatially limited size placed in the desired position.

Since the energy of a harmonic oscillator is
\begin{equation}
    \hat{E} = {1 \over 2} \left ( \hat{p}^2 + \hat{q}^2 \right ),
\end{equation}
it is easily shown that
\begin{equation}
    \mathrm{Re} [(p^2)_w(q)] = 2 \mathrm{Re} [E_w(q)] - q^2,
\end{equation}
where $E_w(q)$ is the weak value of energy postselected on
position. This suggests an alternative measurement strategy of
$\mathrm{Re}[(p^2)_w(q)]$ by performing a weak measurement of
energy postselected on the quadrature $\hat{q}$, and subsequently
subtracting the squared quadrature.

In conclusion, we have investigated a general class of weak
measurements where the state of the pointer could be either pure
or mixed. We have also investigated classical weak measurements.
We have demonstrated that weak measurements will reveal
nonclassical properties of coherent states. We demonstrated that
weak values are conditional moments of the Margenau-Hill
distribution, and that nonclassicality of coherent states is
related to negativity of the Margenau-Hill distribution.

\acknowledgements

The author acknowledges constructive criticism from one of the
referees which led to several improvements of this Letter.

\end{document}